\documentclass[reprint,superscriptaddress,amsmath,amssymb,aps,showkeys
]{revtex4-2}

\usepackage{graphicx,xcolor}
\usepackage{dcolumn}
\usepackage{bm}

\newcommand{\av}[1]{\langle #1 \rangle}

\begin{document}

\preprint{APS/123-QED}

\title{Classical and Quantum Elliptical Billiards: Mixed Phase Space and Short Correlations in Singlets and Doublets}

\normalsize

\author{T. Ara\'ujo Lima}
\thanks{Corresponding author: tiago.araujol@ufrpe.br}
\affiliation{Departamento de F\'isica, Universidade Federal Rural de Pernambuco, Recife, PE 52171-900, Brazil}

\author{R. B. do Carmo}
\thanks{ricardo.carmo@ifal.edu.br}
\affiliation{Instituto Federal de Alagoas, Piranhas, AL 57460-000, Brazil}

\date{\today}

\begin{abstract}

Billiards are flat cavities where a particle is free to move between elastic collisions with the boundary. In chaos theory these systems are simple prototypes, their conservative dynamics of a billiard may vary from regular to chaotic, depending only on the border. The results reported here seek to shed light on the quantization of classically chaotic systems. We present numerical results on classical and quantum properties in two bi-parametric families of Billiards, Elliptical Stadium Billiard (ESB) and Elliptical-$C_3$ Billiards (E-$C_3$B). Both are elliptical perturbations of chaotic billiards with originally circular sectors on their borders. Our numerical calculations show evidence that the elliptical families can present a mixed classical phase space, identified by a parameter $\rho_\text{c} < 1$, which we use to guide our analysis of quantum spectra. We explored the short correlations through nearest neighbor spacing distribution $p(s)$, which showed that in the mixed region of the classical phase space, $p(s)$ is well described by the Berry-Robnik-Brody (BRB) distributions for the ESB. In agreement with the expected from the so-called ergodic parameter $\alpha = t_\text{H}/t_\text{T}$, the ratio between the Heisenberg time and the classical diffusive-like transport time signals the possibility of quantum dynamical localization when $\alpha < 1$. For the E-$C_3$B family, the eigenstates can be split into singlets and doublets. BRB describes $p(s)$ for singlets as the previous family in the mixed region. However, the $p(s)$ for doublets are described by new distributions recently introduced in the literature but only tested in a few cases for $\rho_\text{c} < 1$. We observed that as $\rho_\text{c}$ decreases, the $p(s)$'s tend to move away simultaneously from the GOE (singlets) and GUE (doublets) distributions.


\end{abstract}

\maketitle


\section{Introduction}
\label{intro}

The idea that molecules may be behind Thermodynamics (grounded in Statistical Mechanics) was one of the tremendous scientific advances of the 19th century. In particular, these particles, constituents of gases, are associated with the concept of ergodicity, then called \emph{molecular chaos}. The word ergodic came from the Greek \emph{ergon} (work) and \emph{odos} (trajectory) and was used by Boltzmann to represent the hypothetical visit to all points of the phase space by a particle of that gas with random microscopic dynamic behavior. The introduction of the probability in theory that came to be called Statistical Mechanics of Equilibrium passed by a long probationary regime, with more convincing results occurring only in the first decades of the 20th century \cite{dor:1999}. The so-called Ergodic Hypothesis only gained the rigor of a theorem with the work of the Russian mathematician Y. Sinai in the 60s-70s for an ideal gas of only two particles \cite{sin:1970}. A system is chaotic if two neighboring trajectories in the phase space separate exponentially. Suppose the distance in phase space between such trajectories is proportional to $e^{\lambda t}$. The $\lambda$ parameter is called the Lyapunov exponent. In reality, $\lambda$ represents the greatest of Lyapunov's exponents. Therefore, the existence of at least one positive Lyapunov exponent characterizes a chaotic system \cite{ott:2002}. Billiards systems are prototypes in the study of chaos and describe the free movement of a point particle in a closed domain $\Omega$ with elastic reflections on the boundary $\partial \Omega$ of the domain. The nature of this conservative dynamical system depends exclusively on the shape of the border $\partial \Omega $, varying from entirely regular (i.e., ellipses and annular concentric regions) to completely chaotic (i.e., Sinai billiard). Without loss of generality, we consider that the particle has mass $m=1$ and velocity of module $| \bm{v} | $ $=1$. A discrete dynamics well describes this 2-dimensional motion in time on variables $(\ell,\phi)$, the fraction of perimeter of $\partial \Omega$, and the incidence angle where a collision happens parametrizes the discrete-time generally \cite{che:2006}. A primordial example that deserves to be mentioned here is the Bunimovich stadium. This billiard can present $\lambda>0$. Its shape consists of two semicircles joined by two finite-size segments $2t$, forming a stadium. It is chaotic for any $t>0$. In a pure circular billiard, collisions keep the angular momentum in relation to its center (focus) constant. The Bunimovich stadium does not present this property in its dynamics, known as a defocusing \cite{bun:1974}. It is hugely relevant to this work because the Elliptical Stadium Billiard is a perturbation of it, resulting in classical dynamics with mixed phase space.

Quantum mechanics has been one of the best-tested physical theories since its emergence. The theory makes excellent predictions not only for the atom of hydrogen, which is classically integrable, as well as the helium atom, which is classically not integrable. Nothing is more natural than whether there is an effect analogous to chaos in quantum mechanics. 
The term \emph{quantum chaos} is generally understood as studying the quantum behavior of classically chaotic systems \cite{sto:2000}. One commonly used means of studying these systems is to statistically characterize spectral properties in the semiclassical regime and compare them with results from the random matrices theory \cite{meh:2004}.

In billiards, obtaining the energy spectrum is an essential step for analysis. The problem is to solve the time-independent Schr\"odinger equation with null potential in the planar region $\Omega$ with Dirichlet boundary conditions at $\partial \Omega$:
\begin{equation}
 \left\{ \begin{array}{l}
  \nabla^2 \varphi_n(\bm{r}) = - k_n^2 \varphi(\bm{r}), \quad \bm{r} \in \Omega\\
  \varphi_n(\bm{r})=0, \quad \bm{r} \in \partial \Omega, \end{array}\right.
 \label{eq:helm}
\end{equation}
expression is also known as the Helmholtz Equation \cite{has:1999}. Where $k_n^2=2mE_n/\hslash^2$. In order to characterize universality, one must first unfold the energy spectrum $\{ E_n \}$ so that a unit means ($\av{s_n} = 1$) nearest neighbor spacing (nns) $s_n = E_{n+1}-E_n$ is obtained. This approach became relevant after two important conjectures. Namely, the Berry-Tabor (BT) conjecture \cite{ber:1977} and the Bohigas-Giannoni-Schmit (BGS) conjecture \cite{boh:1984}. The BT conjecture states that, in the semiclassical limit, the statistical properties of the energy spectrum of a classically integrable system must correspond to the prediction of uncorrelated randomly distributed energy levels. As a result, the semiclassical nns distribution $p(s)$ must obey Poisson:
\begin{equation}
 p_\text{P}(s) = \exp (-s).
 \label{eq:poi}
\end{equation}
On the other hand, according to the BGS conjecture, in the case of a classically chaotic system, the spectral properties must follow the universal statistics of the eigenvalues of Gaussian random matrices \cite{meh:2004}. Several recent works have improved the turnover of the BGS conjecture in a theorem \cite{mul:2004,mul:2005,heu:2007,mul:2009}. These proofs still have controversies and limitations pointed out by some authors \cite{ull:2016,shn:2020}. The terminology "BGS conjecture" fits the current article for quantized billiards. More recently, in \cite{loz:2022}, the conjecture was extended to purely ergodic systems. If one disregards spin, in the presence (absence) of time-reversal symmetry, $p(s)$ must correspond to that of the GOE, \emph{Gaussian Orthogonal Ensemble} (GUE, \emph{Gaussian Unitary Ensemble}):
\begin{equation}
 \left\{ \begin{array}{l}
  p_\text{GOE}(s) = (\pi/2) s \exp (-\pi s^2/4),\\
  p_\text{GUE}(s) = (32/\pi^2) s^2 \exp (-4 s^2/\pi).\end{array}\right.
 \label{eq:goegue}
\end{equation}
Based on these assumptions, Leyvraz, Schmit, and Seligman (LSS) \cite{ley:1996} predicted and tested numerically that chaotic billiards with only a three-fold ($C_3$) symmetry (without reflection symmetry) have doublets with spectral statistics of the GUE type, although billiards are by time reversal. LSS considered a billiard consisting of three straight segments of an equilateral triangle with rounded corners by two circumferences of different radii, here called Circular-$C_3$ Billiard (C-$C_3$B). In particular, LSS showed results for a double ratio between the radii where there is a satisfactory agreement for $p(s)$ with the GUE statistics, for a total of approximately 800 doublets. Later, C. Dembowski \emph{et al.} used microwave billiards with $C_3$ symmetry to check experimentally the result predicted by LSS. Besides, they showed that singlets follow GOE \cite{dem:2003}.

The BT \cite{cas:1985}, BGS \cite{rob:1984,lop:2006,bat:2019} and LSS conjectures have been investigated in the literature, but there have been comparatively fewer studies on the LSS findings \cite{die:2005,men:2007,tek:2020,ara:2021}. Until now, little has been said about the situations of $C_3$ symmetric billiards with mixed classical phase space, where chaotic sea and stable KAM-islands coexist. Here, we propose to shed light on the quantum properties of billiards with mixed classical phase space. For this, we perform numerical calculations on the energy spectra of two bi-parametric families of billiards with elliptical sectors on their boundaries and analyze the short correlations. The first one is the Elliptical Stadium Billiard (ESB), a perturbation of the Bunimovich Stadium, whose mixed classical phase space was studied in \cite{can:1998,ara:2015}. In sequence, we introduce a perturbation of the C-$C_3$B, replacing the circumferences with ellipses. Recently, billiards with elliptical borders have been studied in other contexts, i.e., in singular potentials \cite{die:2022}, in relativistic limits \cite{yu:2022}, and flows that move around chaotic cores \cite{bun:2022}. We start the analysis by presenting the billiards, discussing their classical dynamics, showing some mixed phase spaces, and calculating the fraction of the chaotic sea on these phase spaces. Finally, we follow with the quantized billiards' spectral properties, investigating the nns distribution $p(s)$ with formulas for intermediate quantum statistics derived for the doublets recently \cite{ara:2021}.

\section{The bi-parametric billiards families and Classical Dynamics}
\label{billiards}

The billiards systems studied in this work belong to two bi-parametric families, the Elliptical Stadium Billiards (ESB) and Elliptical-$C_3$ Billiards (E-$C_3$B). The first one consists of a perturbation of the Bunimovich Stadium. It comprises two half-ellipses (major semi-axis $a$ and minor semi-axis $1$) that bracket a rectangular sector of thickness $2t$ and height $2$. \cite{can:1998} showed that in the region $a \in (1,\sqrt{2})$ and $t \in (0,\infty)$ are possible to find chaotic dynamics or a mixed phase space depending on the parameters. In \cite{ara:2015} is presented a critical behavior of the billiard dynamics near a transition curve, $t(a)=\sqrt{a^2-1}$ for the interval $a \in (1,\sqrt{4/3})$. Based on these previous works, we focus our analysis on this last interval and $t \in (0,1/\sqrt{3})$. The E-$C_3$B is based on C-$C_3$B, but ellipses instead of circumferences curve the corners. The larger (smaller) ellipse has $A_\text{e}$ $(a_\text{e})$ and $B_\text{e}$ $(b_\text{e})$ semi-axes. In all cases described here, the relations $A_\text{e}=2a_\text{e}$ and $B_\text{e}=2b_\text{e}$ are maintained, with $a_\text{e}$ and $b_\text{e}$ in the range $(0,\sqrt{3}/6)$. The LSS billiard is reproduced with $a_\text{e}=b_\text{e}=\sqrt{3}/12$. Here, by our knowledge, we present for the first time a perturbation on the C-$C_3$B resulting in a system that shows a mixed phase space.

\begin{figure}[!htpb]
 \centering
 \includegraphics[width=0.9\columnwidth]{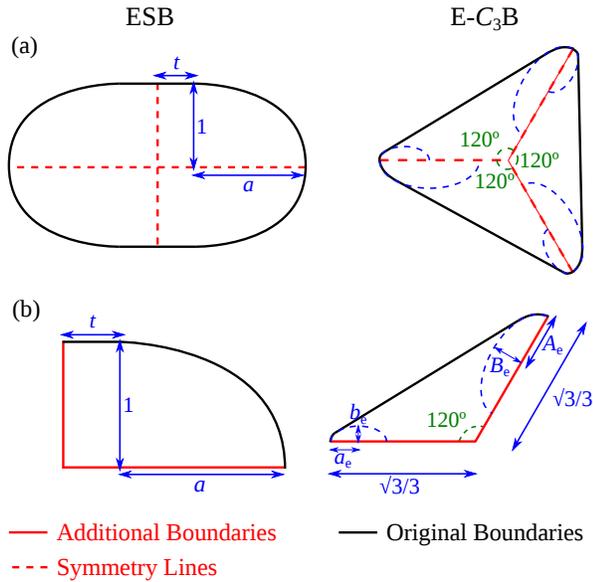}
 \caption{(a) Original Boundaries of Elliptical Stadium Billiard and Elliptical-$C_3$ Billiard. For ESB, the symmetry lines are referent to reflections on vertical and horizontal axes. While E-$C_3$B are referent to $120^{\circ}$ rotational axes. (b) Fundamental Domains of ESB and E-$C_3$B. The symmetry lines are replaced by additional boundaries forming the planar region where we analyze these billiards' classical dynamics.}
 \label{fig:billiards}
\end{figure}

A Fundamental Domain (FD) is a neighborhood in $\Omega$ that contains only one image for any point in the system. Besides the boundary of the $\partial \Omega$, there are additional boundaries between adjacent FDs, which are the symmetry lines. Classically, billiard dynamics can always be reduced to a FD by assuming specular reflections at the symmetry lines \cite{cvi:1993,li:2020}. For this, we use the FD of each billiard in our calculations on the classical dynamics. In Fig. \ref{fig:billiards}, we graph billiards in families indicating the parameters and their respective FDs.

\begin{figure}[!htpb]
 \centering
 \includegraphics[width=0.95\columnwidth]{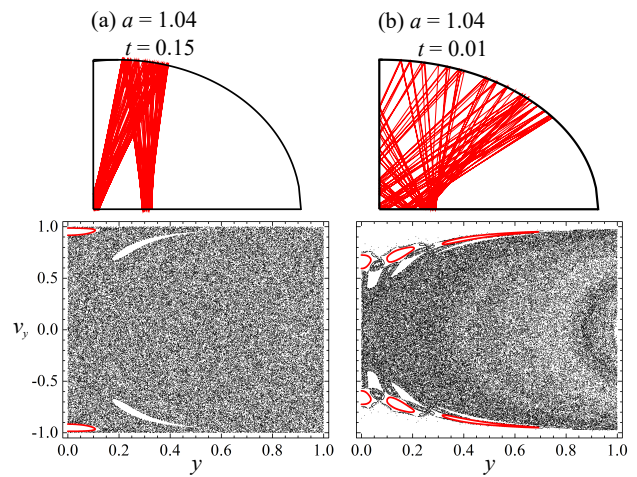}
 \caption{Upper Panels: ESB boundaries for some values of parameters $(a,t)$ with stable trajectories in red. Lower Panels: corresponding phase portraits for $10^5$ collisions with the vertical boundary from the IC $(y_0,v_{y0})=(0.5,0.0)$ (small gray dots). The red plots correspond to the trajectories in the upper panels. The mixed phase spaces present $\rho^{(\text{a})}_\text{c} = 0.991884$ and $\rho^{(\text{b})}_\text{c} = 0.857184$.}
 \label{fig:PS_ESB}
\end{figure}

\begin{figure}[!htpb]
 \centering
 \includegraphics[width=0.95\columnwidth]{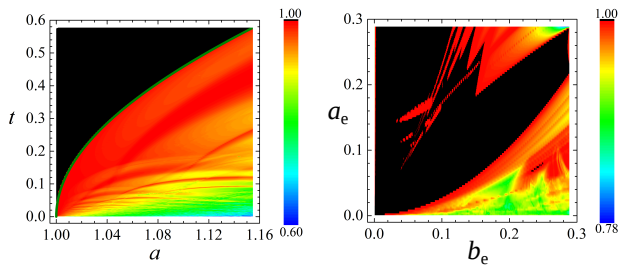}
 \caption{Left Panel: diagram of the chaotic fraction of the phase space $\rho_\text{c}$ for the ESB. The tiny green line is the critical line $t(a)=\sqrt{a^2-1}$ studied in \cite{can:1998,ara:2015}. Right Panel: same diagram for the E-$C_3$B showing a distinguished phase space behavior depending on the parameters. The ergodic property $(\rho_\text{c} = 1)$ is numerically guaranteed in black regions. These maps will guide us in exploring quantum properties, where these values will be relevant parameters to our analysis.}
 \label{fig:diagrams}
\end{figure}

The global dynamical properties of the ESB with unit mass and velocity may be characterized through collisions of orbits with the vertical side of its FD shown in Fig. \ref{fig:billiards}. An additional part of the boundary dictates this edge and does not change with the variation of parameters. The reduced phase space is then a rectangle defined by the vertical position $y$, where a collision occurs at discrete time $n$, and the tangent component of the velocity in a collision, $v_y$, with $0 < y < 1$ and $-1 < v_y < 1$. The small gray dots in Fig. \ref{fig:PS_ESB} show the phase plane for some values of parameters $(a,t)$ after $n=10^5$ collisions from the initial conditions (ICs), clearly exhibiting a mixed (regular-irregular) characteristic. We plot one example of a stable trajectory in red for each one. Quantitative characterization of these mixed-phase spaces can be made through the chaotic (regular) fraction $\rho_\text{c}$ ($\rho_\text{r}$) of each phase portrait with $\rho_\text{c} + \rho_\text{r} = 1$ and $0 \leqslant \rho_\text{c} \leqslant 1$. The phase plane is partitioned into $N_\text{c}$ small disjoint cells to measure these quantities \cite{rob:1997,cas:1999,ara:2013,ara:2015,ara:2021}. For a given orbit, let $N(n)$ be the number of different cells in the phase space, which are visited up to $n$ impacts in the cross-section. The relative measure $r(n)$ is defined as the fraction of visited cells averaged over a set of ICs, i.e., $r(n) = \av{N(n)}/N_\text{c}$. So the chaotic fraction of the phase space is obtained via
\begin{equation}
 \rho_\text{c} = \lim_{n \to \infty} \lim_{N_\text{c} \to \infty}  r(n),
 \label{eq:rn_rhoc}
\end{equation}
for ICs in the chaotic sea. In our numerical approach, we consider $N_\text{c} = 10^6$, $n = 2\cdot10^7$, and averages in 20 random ICs. For the billiards with mixed phase space in Fig. \ref{fig:PS_ESB}, $\rho^{(\text{a})}_\text{c} = 0.991884$ and $\rho^{(\text{b})}_\text{c} = 0.857184$. The left panel of Fig. \ref{fig:diagrams} shows a numerical diagram of $\rho_\text{c}$. The ergodic property $(\rho_\text{c} = 1)$ is numerically guaranteed in black regions. This diagram also supports previous works \cite{can:1998,ara:2015}, where a critical transition from a mixed phase space to a fully ergodic was found to cross a critical line $t(a)=\sqrt{a^2-1}$.

\begin{figure}[!htpb]
 \centering
 \includegraphics[width=0.95\columnwidth]{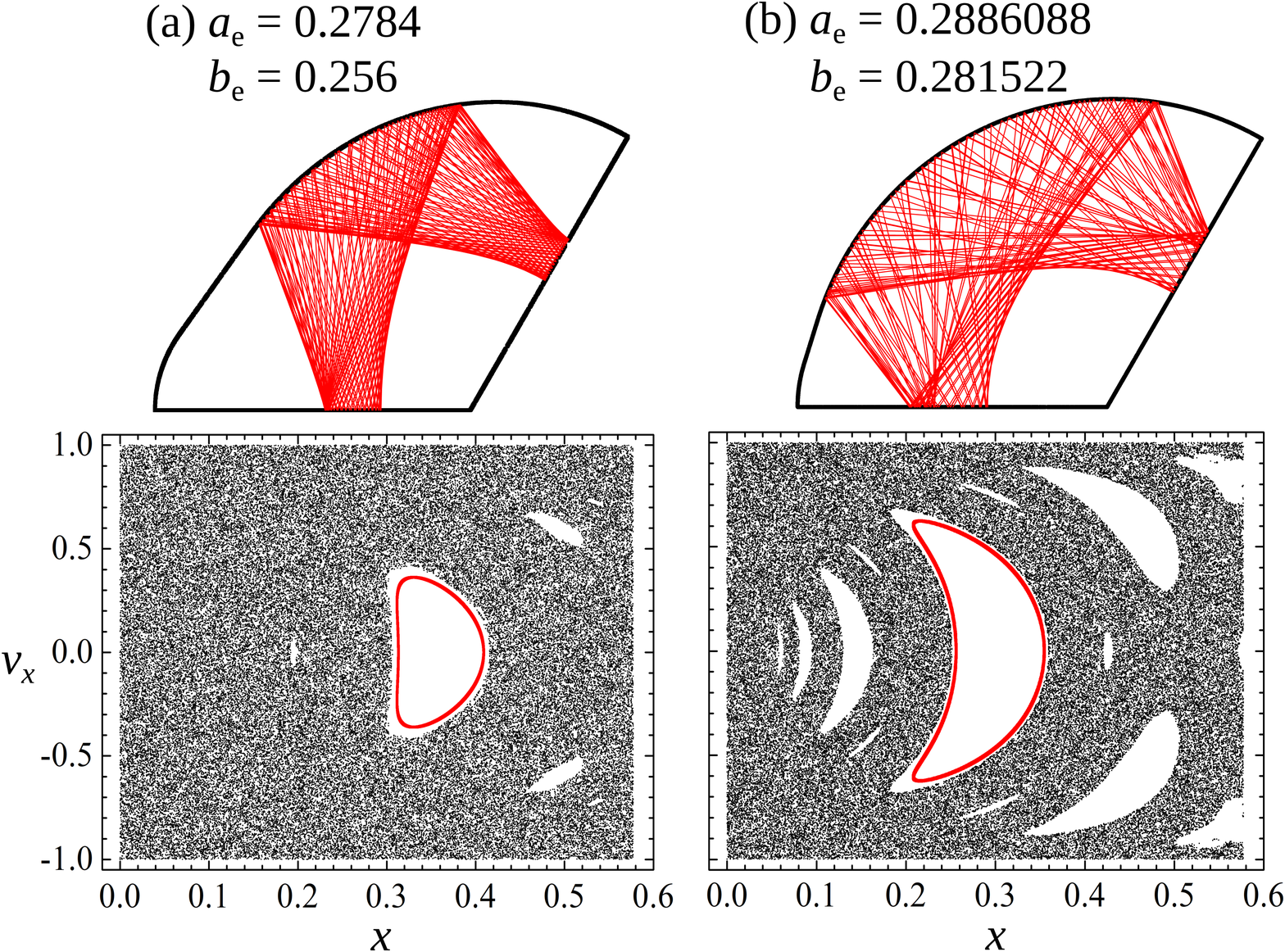}
 \caption{Upper Panels: E-$C_3$B boundaries for some values of parameters $(a_\text{e},b_\text{e})$ with stable trajectories in red. Lower Panels: corresponding phase portraits for $2\cdot10^7$ collisions with the horizontal boundary from the IC $(x_0,v_{x0}) = (0.5,0.0)$ (small gray dots). The red plots correspond to the trajectories in the upper panels. The mixed phase spaces present $\rho^{(\text{a})}_\text{c} = 0.935152$ and $\rho^{(\text{b})}_\text{c} = 0.800792$.}
 \label{fig:PS_EC3B}
\end{figure}

The E-$C_3$B's classical dynamical properties will be studied in the same way but are characterized through the collisions of the orbits with the horizontal side of its FD shown in Fig. \ref{fig:billiards}, which does not change with the variation of parameters. The reduced phase space is then a rectangle defined by the horizontal position $x$, and the tangent component of the velocity in a collision, $v_x$, with $0 < x < \sqrt{3}/3$ and $-1 < v_x < 1$. The small gray dots in Fig. \ref{fig:PS_EC3B} show the phase plane for some values of parameters $(a_\text{e},b_\text{e})$ after $n=2\cdot10^7$ collisions from the ICs, clearly exhibiting mixed (regular-irregular) characteristic. The values of chaotic fraction are $\rho^{(\text{a})}_\text{c} = 0.935152$ and $\rho^{(\text{b})}_\text{c} = 0.800792$. The right panel of Fig. \ref{fig:diagrams} shows a numerical diagram of $\rho_\text{c}$. The ergodic property $(\rho_\text{c} = 1)$ is numerically guaranteed in black regions. This map will guide us in exploring quantum properties described in the next section, where these values will be relevant parameters to our analysis.

\section{Quantization and Eigenvalues Short Correlations}
\label{quantum}

All Energy spectra $\{ E_n \}$ of eq. (\ref{eq:helm}) were calculated with an algorithm based on the scaling method introduced by E. Vergini and M. Saraceno (VS) in \cite{ver:1995}. This approach allows us to access high-lying energy eigenvalues that have been unfolded to obtain a unit mean spacing ($\av{s_n} = 1$) for each billiard. Our results are based on sets of approximately 70,000 eigenvalues for a given pair of parameters. According to \cite{sto:2000}, there is possibly no more intensely studied spectral statistics more than $p(s)$, the density of probability of finding two levels nearest neighbor spaced by $s$.

\subsection{The Singlets Case}
\label{singlets}

Initially, we focused on results for ESB. Some proposes have been made to describe these distributions for systems whose present mixed-phase space on its classical counterpart. Here we focus on two of them. They result in intermediate formulas between Poisson and GOE statistics through parameters variation. Firstly, we cite the purely phenomenologic approach by Brody \cite{bro:1973}, where an exponent $\nu$ is gradually varied to obtain a smooth change between the integrable ($\nu=0$) and chaotic ($\nu=1$) cases:
\begin{equation}
 p_\text{B}(s) = a_\nu (\nu+1)s^\nu \exp \left [ -a_\nu s^{(\nu+1)} \right ],
 \label{eq:brody}
\end{equation}
where $a_\nu = \left [ \Gamma \left ( \frac{\nu+2}{\nu+1} \right ) \right ]^{\nu+1}$ and $\Gamma(x)$ is the Gamma function. The second distribution cited here is the Berry-Robnik-Brody (BRB), a proposal that takes under consideration the chaotic (regular) fraction of the classical phase space $\rho_\text{c}$ ($\rho_\text{r}$) \cite{bat:2010}:
\begin{multline}
 p_\text{BRB}(s)= \\
 \exp(-\rho_\text{r} s)\left \{ \frac{\rho_\text{r}^2}{(\beta+1)\Gamma \left (\frac{\beta+2}{\beta+1} \right )} Q \left [ \frac{1}{\beta+1};a_\beta (\rho_\text{c} s)^{\beta+1} \right ] + \right. \\
 \left. [ 2 \rho_\text{r} \rho_\text{c} + (\beta+1) a_\beta \rho_\text{c}^{\beta+2}s^\beta ] \exp[-a_\beta(\rho_\text{c} s)^{\beta+1}] \right \}.
 \label{eq:brb}
\end{multline}
As in the Brody distribution, $a_\beta = \left [ \Gamma \left ( \frac{\beta+2}{\beta+1} \right ) \right ]^{\beta+1}$ and $Q(\kappa;x)$ is the Incomplete Gamma function. This distribution can go through other distributions varying the free parameters $\rho_\text{c}$ and $\beta$. For $\beta=0$, $p_\text{BRB}(s)=p_\text{P}(s)$ and for $\beta=1$ it recovers the distribution of Berry-Robnik (BR) \cite{ber:1984}. If $\rho_\text{c}=0$, $p_\text{BRB}(s)=p_\text{P}(s)$ again, while for $\rho_\text{c}=1$, $p_\text{BRB}=p_\text{B}(s)$.

\begin{figure*}[!htpb]
 \centering
 \includegraphics[width=0.95\textwidth]{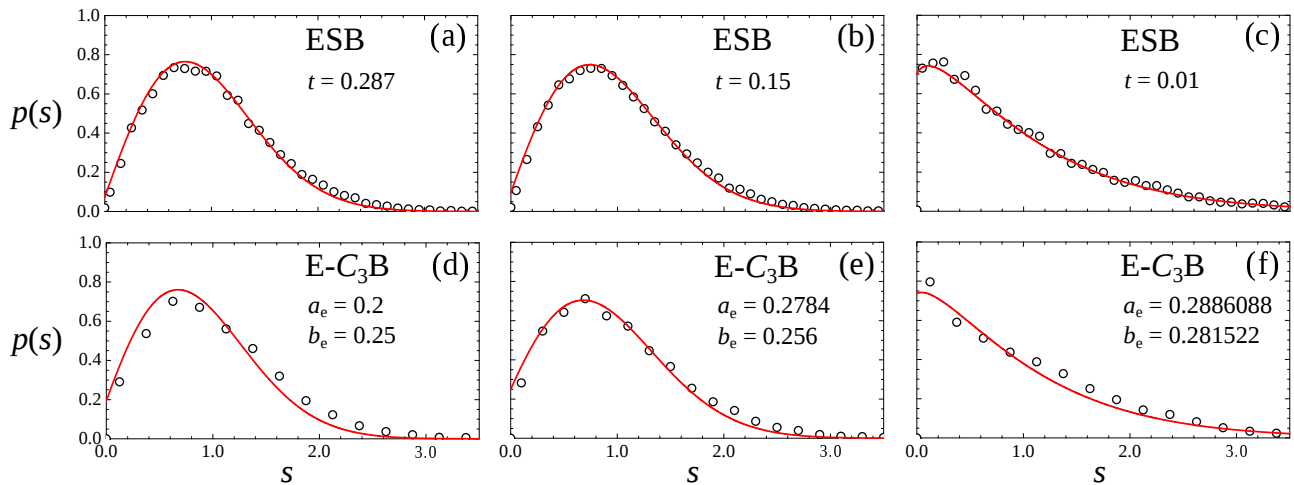}
 \caption{Representative results for BRB distributions fits for $p(s)$. Upper Panels: results on ESB with $a=1.04$ and some values of $t$. The chaotic case $t=0.287$ $(\rho_\text{c} = 1)$ presents $\beta = 1.000 \pm 0.020$, the GOE distribution. The mixed cases $t=0.15$ and $t=0.01$ ($0 < \rho_\text{c} < 1$) present $\beta = 0.978 \pm 0.018$ and $\beta = 0.191 \pm 0.014$ respectively, intermediate distributions between Poisson and GOE. Lower Panels: results on E-$C_3$B with some values of $(a_\text{e},b_\text{e})$. The chaotic case, $(a_\text{e},b_\text{e})=(0.2,0.25)$ ($\rho_\text{c} = 1$) presents $\beta = 1.000 \pm 0.097$, the GOE distribution. The mixed cases, $(a_\text{e},b_\text{e})=(0.2784,0.256)$ and $(a_\text{e},b_\text{e})=(0.2886088,0.281522)$ ($0 < \rho_\text{c} < 1$) present $\beta = 0.999 \pm 0.057$ and $\beta = 0.203 \pm 0.073$ respectively, in the range of intermediate distributions between Poisson and GOE. The fits with the Brody formula and BRB distribution are indistinguishable in both billiards families.}
 \label{fig:p(s)}
\end{figure*}

The nns for ESB were previously studied in \cite{lop:2006} with around 3,000 eigenvalues of eq. (\ref{eq:helm}). We use the VS method to obtain around 65,000 eigenvalues beyond the first 5,000. The BRB distribution can fit all $p(s)$ obtained for all parameters tested on ESB. We have two independent parameters for this distribution, $\rho_\text{c}$, and $\beta$. However, we fixed $\rho_\text{c}$ at the value obtained in the diagram of Fig. \ref{fig:diagrams}. The upper panels of Fig. \ref{fig:p(s)} shows representative results. The chaotic case presents $\beta = 1.000 \pm 0.020$, the GOE distribution. The mixed ($0 < \rho_\text{c} < 1$) present $\beta = 0.978 \pm 0.018$ and $\beta = 0.191 \pm 0.014$, intermediate distributions between Poisson and GOE. These results go in the direction of the quantum localization, previously studied in other billiards systems \cite{bat:2013,loz:2021} and discussed next.

\begin{figure}[!htpb]
 \centering
 \includegraphics[width=0.9\columnwidth]{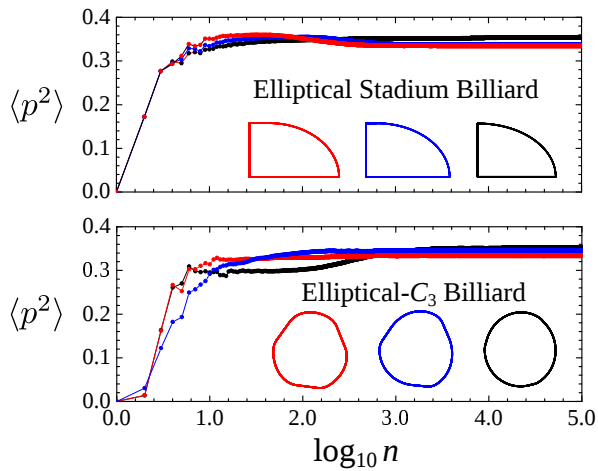}
 \caption{Calculated mean square of the momentum as a function of the discrete time $n$ in a monolog scale (number of collisions of the particle off the billiard boundary). Lines are guides for the eyes. Upper panel: results for members of the ESB family with $a=1.04$. The red dots are for $t=0.287$ and present saturation at $N_\text{T} \simeq 7.10^2$. Blue dots are for $t=0.15$, and saturation at $N_\text{T} \simeq 2.10^3$, and black dots are for $t=0.01$ presenting $N_\text{T} \simeq 6.10^2$. Lower Panel: same calculations for the E-$C_3$B, the red dots are for $(a_\text{e},b_\text{e})=(0.2,0.25)$ and present saturation at $N_\text{T} \simeq 3.10^2$. Blue dots are for $(a_\text{e},b_\text{e})=(0.2784,0.256)$ and saturation at $N_\text{T} \simeq 7.10^2$, and, black dots are for $(a_\text{e},b_\text{e})=(0.2886088,0.281522)$ presenting $N_\text{T} \simeq 2.10^3$.}
 \label{fig:NT}
\end{figure}

Quantum dynamical localization corresponds to a peculiar quantum distribution of the linear or angular momentum peaked at zero, with walls that decay exponentially, differently from the classical results, which predicts, for a chaotic or disordered system, a diffusive transport \cite{bor:1996}. The phenomenon can be reviewed in \cite{pro:2000}. An interesting feature of the quantum dynamical localization is that it allows us to estimate the conditions under which the comparison with
the standard random matrix theory is adequate or, in other words, whether an energy eigenvalues data set belongs to the deep semiclassical regime. We follow closely \cite{bat:2013} in the short description below. The key idea is to express the ergodic parameter $\alpha = t_\text{H} / t_\text{T}$, where $t_\text{H}$ is the (quantum) Heisenberg time, and $t_\text{T}$ is the (classical) transport time, in terms of accessible magnitudes, such as the (quantum) energy $E$ and the (classical) number of collisions off the billiard border, $N_\text{T}$. From \cite{bat:2013} the ratio is expressed as
\begin{equation}
 \alpha = \frac{k \mathcal{L}}{\pi N_\text{T}},
 \label{eq:alpha}
\end{equation}
where $\mathcal{L}$ is the perimeter of the boundary and $k^2 \sim E$. The condition for quantum dynamical localization in a given energy spectrum, $\alpha \leqslant 1$, can then be written as $k \leqslant k_\text{c} = \pi N_\text{T}/\mathcal{L}$. To estimate $N_\text{T}$, we consider an ensemble of orbits initially directed perpendicularly to $\partial \Omega$ and follow its random spreading as a function of the discrete time $n$. The symbols in Fig. \ref{fig:NT} illustrate the results for the mean square momentum $\av{p^2}$ as a function of $n$ in a monolog scale (averaged in sets of $10^3$ randomly chosen ICs) for members of two billiards family. Saturation of $\av{p^2}$ occurs at different times $N_\text{T}$ depending on parameters. For the ESB family, all calculated spectra have $k_{max} \lesssim k_\text{c}$ as the largest eigenvalue, equivalent to the 70,000th level at least. These facts are in agreement with the intermediate statistics well fitted with eq. (\ref{eq:brb}) as in \cite{bat:2010,bat:2013,bat:2019,ara:2021}. The same occurs for the singlets in the E-$C_3$B family, where the condition $k_{max} \lesssim k_\text{c}$ is equivalent to the 70,000th level. The representative results are in the lower panels of Fig. \ref{fig:p(s)}. The chaotic case presents $\beta = 1.000 \pm 0.097$, the GOE distribution. The mixed ($0 < \rho_\text{c} < 1$) present $\beta = 0.999 \pm 0.057$ and $\beta = 0.203 \pm 0.073$, in the range of intermediate distributions between Poisson and GOE. In the next section, we discuss the doublets subspace.

\subsection{The Doublets Case}
\label{doublets}

\begin{figure}[!htpb]
 \centering
 \includegraphics[width=0.9\columnwidth]{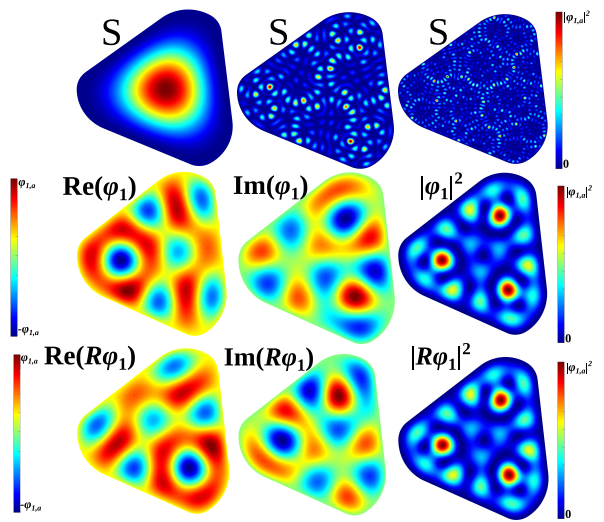}
 \caption{Top panels: Density plots of squared eigenfunctions corresponding to singlet states in the LSS billiard, exhibiting the underlying $C_3$ symmetry. In the color scale, $|\varphi_{1,a}|^2$ is the maximum probability in each case. Middle panels: Real and imaginary parts of a member $\varphi_1$ of a doublet. In the color scale, $\pm \varphi_{1,a}$ is the minimum and maximum of the wave function. The probability density recovers the $C_3$ symmetry (right panels). Bottom panels: Same state in the middle under the application of the rotation operator $R(2\pi/3)$.}
 \label{fig:singdoub}
\end{figure}

Consider a classically chaotic system with time-reversal (TR) invariance and a point-group (PG) symmetry. If the TR and the PG operations do not commute, non-self-conjugate invariant subspaces of the PG must exhibit GUE spectral fluctuations instead of GOE ones \cite{ley:1996}. For example, consider a billiard in the $xy$ plane with the $C_3$ symmetry. Such a billiard has eigenfunctions $\varphi_m$ $(m = -1, 0, +1)$, such that $\varphi_0$ is symmetric and repeats itself after a rotation of $2\pi/3$ about the symmetry axis, whereas $\varphi_{\pm 1}$ will be repeated only after three consecutive rotations of $2\pi/3$. In other words, if $R(2\pi/3)$ is the rotation operator for an angle of $2\pi/3$, one has $R(2\pi/3)\varphi_m = \exp(i\frac{2\pi}{3} m)\varphi_m$. Let $\Theta$ be the time reversal operator. $\Theta$ is an antiunitary operator that commutes with the Hamiltonian $H$, which has eigenvalue $E_m$, \emph{i.e.}, $H\varphi_m = E_m\varphi_m$. It follows that $H\Theta\varphi_m = \Theta H \varphi_m = E_m\Theta\varphi_m$ ($\Theta \varphi_m$ is also an eigenfunction of $H$ with the same eigenvalue $E_m$). Are $\varphi_m$ and $\Theta \varphi_m$ the same eigenstate? For this subspace one may write $\Theta \varphi_m = (-1)^m \varphi_{-m}$. Thus, $\Theta \varphi_0 = \varphi_0$, \emph{i.e.}, $\varphi_0$ is a singlet. The top panels in Fig. \ref{fig:singdoub} show cases of the probability density $|varphi_0|^2$. On the other hand, $\varphi_1$ and $\varphi_{-1}$ must correspond to distinct states. One refers to this doublet state as a Kramers degeneracy. The middle panels in Fig. \ref{fig:singdoub} show the real and imaginary parts of the member $\varphi_1$ of a doublet, say $(\varphi_1,\varphi_{-1})$, in the same billiard. The probability density $|\varphi_1|^2$ recovers the $C_3$ symmetry (rightmost middle panel in Fig. \ref{fig:singdoub}). The bottom panels in Fig. 7 show the same state under the application under rotation operator $R(2\pi/3)$. A complex conjugation of the shown state obtains the other member $\varphi_{-1}$ of the doublet. Since these degenerate states are not TR invariant, they must follow the GUE of random matrices, providing the billiard is classically chaotic, according to the LSS results.

\begin{figure}[!htpb]
 \centering
 \includegraphics[width=0.9\columnwidth]{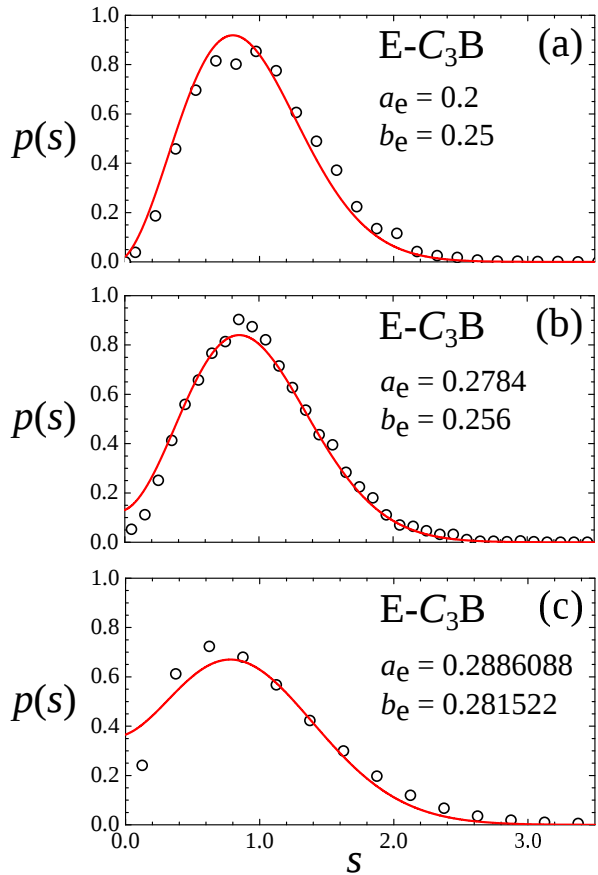}
 \caption{Representative results for BRB-like distributions, eq. (\ref{eq:brb2}), fits for $p(s)$ in doublets subspace for same members of E-$C_3$B family of Fig. \ref{fig:p(s)}. In panel (a), the chaotic case $(a_\text{e},b_\text{e})=(0.2,0.25)$ ($\rho_\text{c} = 1$) presents $\gamma = 0.960 \pm 0.050$, in the range of a GUE distribution. In panels (b) and (c), the mixed cases $(a_\text{e},b_\text{e})=(0.2784,0.256)$ and $(a_\text{e},b_\text{e})=(0.2886088,0.281522)$ ($0 < \rho_\text{c} < 1$) present $\gamma = 1.000 \pm 0.032$ and $\gamma = 1.00 \pm 0.13$ respectively, in the range of intermediate distributions between Poisson and GUE. Fits with Brody-like formula (\ref{eq:brody2}), and BRB-like distribution (\ref{eq:brb2}), are indistinguishable.}
 \label{fig:p(s)_doub}
\end{figure}

For the E-$C_3$B, the degenerate states remain invariant to TR. However, the spectral distribution will be changed for cases where the classical dynamics is not completely chaotic ($\rho_\text{c} < 1$), with a $p(s)$ resultant that deviates from the GUE case. Thus, it is necessary to use new intermediate formulas to study the distribution of doublets in billiards with mixed classical phase space. The following formulas we derived in \cite{ara:2021}. Following the same steps in \cite{bro:1973} led to the eq. (\ref{eq:brody}), a Brody-like formula for the transition between the Poisson and GUE distributions is obtained, namely,
\begin{equation}
p_{\text{B,}2}(s) = (\eta+1)b_{\eta}^2s^{2\eta}\exp \left(-b_{\eta}s^{\eta+1} \right),
\label{eq:brody2}
\end{equation}
where 
\begin{equation}
b_{\eta} = \left[\Gamma\left(\frac{2\eta+1}{\eta+1}\right)\right]^{-(\eta+1)},
\label{eq:beta}
\end{equation}
and $0 \leqslant \eta \leqslant 1$. For $\eta = 0$, $p_{\text{B},2}(s)$ reduces to the Poisson distribution, whereas for $\eta = 1$, the Wigner distribution for the GUE is obtained. In \cite{bat:2010}, the dynamical localization of chaotic eigenstates was taken into account and their coupling with the regular ones through tunneling effects. The so-called BRB distribution previously discussed in sec. \ref{singlets}. Following this, the formula that corresponds to the Poisson $\leftrightarrow$ GUE crossover is
\begin{multline}
p_{\text{BRB,2}}(s) e^{\rho_\text{r} s} = \\
\rho_\text{r} \rho_\text{c} b_{\gamma}^{\frac{1}{\gamma + 1}}\left(2 - \rho_\text{r} s \right) Q\left[ \frac{1+2\gamma}{1+\gamma};b_{\gamma}\left(\rho_\text{c} s \right)^{\gamma +1} \right] + \\
\bigg [ \rho_\text{r}^2\left(1+b_{\gamma}\rho_\text{c}^{\gamma+1}s^{\gamma+1}\right) + \\
(1+\gamma)\left(\rho_2^{\gamma+1}b_{\gamma}s^{\gamma} \right)^2 \bigg ] e^{-b_{\gamma}\left(\rho_\text{c} s \right)^{\gamma + 1}},
\label{eq:brb2}
\end{multline}
where $b_{\gamma}$ is defined as in eq. (\ref{eq:beta}) and $Q(\kappa;x)$ is the incomplete Gamma function. Here, $p_{\text{BRB,2}}(s) = p_{\text{P}}(s)$ if $\rho_\text{r} = 1$ or if $\gamma = 0$, and $p_{\text{BRB,2}}(s) = p_{\text{B,}2}(s)$ if $\rho_\text{r} = 0$. In \cite{ara:2021}, the above formula was widely tested only in the regime of full ergodicity (polygonal cases) and in a single case with $\rho_\text{c} < 1$. Here, we detail a non-polygonal billiards family that produces a wide variability of $\rho_\text{c}$ values. In these cases, $p_{\text{BRB,2}}$ well-fitted distributions of nns for $\rho_\text{c} < 1$ for all investigated cases. The representative results are in Fig. \ref{fig:p(s)_doub}. As in the previous section, the doublets subspace is in the region of the spectrum such that $k \lesssim k_\text{c}$, equivalent to 60,000th level.

\section{Conclusions ans Perspectives}
\label{conc}

This paper presents numerical results on classical dynamics and quantization in two bi-parametric billiard families. The ESB comprises two ellipses of minor semi-axe unitary, major semi-axe $a$, and a rectangular region of length $2t$ \cite{can:1998,ara:2015}. The other family, introduced here as E-$C_3$B, presents the $C_3$ symmetry \cite{ley:1996,men:2007,ara:2021} and is formed by an equilateral triangle with rounded corners by two ellipses with semi-axis $A_\text{e}=2a_\text{e}$ and $B_\text{e}=2b_\text{e}$. First, we investigate the classical dynamics of these billiards where we built detailed diagrams for the chaotic fraction $(\rho_{\text{c}})$ of their phase spaces. After that, we investigated the nns distributions $p(s)$ for these systems, a measure of short correlations. In the asymmetric ESB family, the parameters space region $(a,t)$ where the classical phase space is mixed (regular and chaotic regions coexist), all found statistics present intermediated results between Poisson and GOE distributions. The BRB distribution \cite{bat:2010}, eq. (\ref{eq:brb}), very well fitted all cases. These results perfectly agree with the expected from the ergodic parameter $\alpha$ that signals the possibility of quantum dynamical localization when $\alpha < 1$. All sets of eigenvalues used as data are in a range of energy that satisfies this condition. In the E-$C_3$B family, the eigenstates can be split into singlets and doublets subspaces due the symmetry. The first subspace presents similar results to the previous family, reinforcing the agreement with the expected energy range set with $\alpha < 1$ \cite{loz:2021}. The doublets subspace, whose for the chaotic cases is expected a GUE distribution shows the more relevant result in this work. All found statistics present intermediated results between Poisson and GUE distributions for the parameter space $(a_\text{e},b_\text{e})$ where the classical phase space is mixed. A BRB-like formula \cite{ara:2021}, eq. (\ref{eq:brb2}), well fitted all cases. This formula was tested for $\rho_\text{c} < 1$ and $\alpha < 1$ in just a few cases. Particularly in the E-$C_3$B family, the minimum value of the chaotic fraction of the classical phase space is $\rho_\text{c} \simeq 0.8$. This limitation can be avoided if we set free the conditions $A_\text{e}=2a_\text{e}$ and $B_\text{e}=2b_\text{e}$, used here to follow closer to the C-$C_3$B introduced by LSS. In this perspective, a phase diagram analog to Fig. \ref{fig:diagrams} even more intricate is generated, possible further explorations of eq. (\ref{eq:brb2}).

The parameter $\beta$ in eq. (\ref{eq:brb}) was extensively compared with other localization metrics, including analyses involving Husimi functions, calculations of the entropy localization measure \cite{bat:2013}, and normalized inverse participation ratio \cite{bat:2019}. How the new distribution, eq. (\ref{eq:brb2}), uses the same arguments to include the parameter $\gamma$ is meritorious in a future comparison between this quantity and other localization metrics. Another theme meritorious of investigation is the level statistics in an energy range that $\alpha \gg 1$. The BR formulas are expected to provide a good description of the deep semiclassical regime \cite{ber:1984}, an excellent agreement has been found with numerical experiments in a billiard for which the eigenvalues set is around 1,500,000th level \cite{bat:2013}, an impressive number. The BR-like formula in \cite{ara:2021} should be tested in a range of high energy in the doublets subspace to close the comparisons between the short correlations in the singlets sets and doublets subspace. In addition, our results indicate an intriguing correlation between singlets and doublets spectra for the E-$C_3$B family, producing $p(s)$'s that move away from the GOE and GUE distributions as $\rho_\text{c}$ decreases, thus requiring a further investigation of the observed effect. In this perspective, a range opens up to investigate the correlation of spectra of different subspaces \cite{abu:2009,abu:2014,li:2020,tek:2020,bho:2021} in billiards that present only rotational symmetries greater than three, which will give the possibility of performing other tests with the new formulas (\ref{eq:brody2}) and (\ref{eq:brb2}).

\begin{acknowledgments}
Useful discussions with F. M. de Aguiar and K. Terto are gratefully acknowledged. This work has been supported by the Brazilian Agencies CNPq, CAPES and FACEPE.
\end{acknowledgments}


\bibliography{refs_2023_TAraujoLRBdoCarmo_EllipticalBilliards.bib}

\end{document}